\documentclass[aps,preprint]{revtex4}
\usepackage{amsfonts}
\usepackage{amsmath}
\usepackage{amssymb}
\usepackage{graphicx}

\input{tcilatex}

\begin{document}

\title{Shock waves in a rotating non-Maxwellian viscous dusty plasma}
\author{Zahida Ehsan$^{1,2}$, M. M. Abbasi$^{3},$ Samiran Ghosh$^{4}$, Majid
Khan$^{3}$ and Muddasir Ali $^{5}$}
\affiliation{$^{1}$Space and Plasma Astrophysics Research Group (SPAR), Department of
Physics, COMSATS University Islamabad, Lahore Campus 54000, Pakistan }
\affiliation{$^{2}$National Centre for Physics, Shahdara Valley Road, Islamabad 45320,
Pakistan}
\affiliation{$^{3}$Department of Physics, Quaid-i-Azam University, Islamabad 45320,
Pakistan\\
$^{4}$Department of Physics, JIS University, Kolkata 700 109, India}
\affiliation{$^{5}$School of Natural Sciences (SNS), National University of Sciences and
Technology (NUST), Islamabad 44000, Pakistan}
\email{For correspondence: ehsan.zahida@gmail.com}
\date{\today }

\begin{abstract}
A theoretical model is presented to study characteristics of dust acoustic
shock in a viscous, magnetized and rotating dusty plasma at both fast and
slow time scales. By employing reductive perturbation technique the
nonlinear Zakharov--Kuznetsov (ZK) \ equation has been derived for both
cases when dust is inactive and dynamic (fast and slow time scales). Both
electrons and ions are considered to follow kappa/Cairns distribution. It is
observed that the viscosity in both cases when dust is in background and
active plays as a key role in dissipation for the propagation of acoustic
shock. Magnetic field and rotation are responsible for the dispersive term.
Superthermality has been found to affect significantly on the formation of
shock wave along with viscous nature of plasma.

The present investigation may be beneficial to understanding the rotating
plasma in particular experiments being carried out.
\end{abstract}

\maketitle

\section{Introduction}

Despite a history spanning nearly a century, research into complex (dusty)
plasmas (consists of nanometers to hundreds of micrometers sized solid
particles in a conventional two component plasma) has progressed
significantly in last two decades mainly after the marvellous observation of
dusty plasma crystals by Thomas et al. in 1994 \cite{1}. Also from more than
ten years dusty plasmas under minute gravity conditions have been studied on
board the International Space Station (ISS) under the joint Russian/German
venture of Plasma Kristall (PK), along with PKE-Nefedov, PK-3 Plus and PK-4
2014 on wards\cite{2}.

Other than novel experimental discoveries of dusty plasma crystals, dust
Mach cones \cite{3}, dust acoustic waves, dust voids \cite{4}, etc., notion
of possible existence of `dust atoms and molecules' was also put forward by
Tsintsadze, Murtaza and Ehsan \cite{5}. Authors later reported
crystallization of dust atoms in the localized region of the electromagnetic
wave\cite{6}.

Importance of dusty plasma physics has been manifold, these are omnipresent
in astrophysical environment like comets, interplanetary of interstellar
clouds, the rings of the Giant planets like Saturn etc., whereas discharges
for thin film deposition or etching, dust in tokamak are few other
noteworthy technological applications \cite{7, 8}.\ These are the reasons, a
rich literature exists on the investigation of linear and nonlinear
structures in a dusty plasma for instance propagation of dust-ion acoustic
waves \cite{9}, dust-acoustic (DA) waves \cite{10}, dust lattice \cite{11}
waves, dust Coulomb waves \cite{12}, dust ion-acoustic shock waves \cite{13}
and dust-acoustic shock waves \cite{14, 15}, cusp solitons \cite{16} etc.
Tsytovich, and Angelis have also contributed significantly in developing
kinetic theory of dusty plasmas \cite{17, 18, 19}.

Since the time scales associated to the big sized dust particles (large mass
to charge ratio) are much longer therefore these plasmas can be tracked on
the individual particle level with the naked eye and so provide an excellent
tool for understanding underlying physics of phase transitions and
collective excitations when in solid liquid or gaseous states. It is the
large mass to charge ratio of dust particles that these plasmas considered
bridging key issues from several fields like warm dense matter,
low-temperature physics, surface and solid-state physics etc.

Conveniently and conventionally in the past modelling of plasma systems was
carried out for static frame of reference whereas actual modelling of large
number of problems in astrophysics (for instance rotating magnetic stars,
pulsar/Kerr black-hole magnetospheres) and in lab (such as tokamak) physics
required to be done in non inertial (in particular rotating) frames.
Chandrasekhar was the first to incorporate non-intertial frames \cite{20}
later Lehnert \cite{21} and Hide \cite{22} also contributed to it. In these
pathbreaking studies it was reported that the tiny force resulted from
rotation (via Coriolis force) has an effective role to play in the plasma
astrophysics and in other cosmic phenomena.

Observations show that the rotating flows of magnetized plasmas are not
uncommon in solar physics, it is for the reason, linear wave propagation has
been studied to show the interaction of the Coriolis force in an ideal lower
ionosphere \cite{23}. Also to understand sunspot development, the star cycle
and the structure of rotating stars magnetospheres etc. non-inertial frames
are inevitable \cite{25}. Coriolis force can also create effective
magnetic-field when the ionized fluid rotates, it for the reason that
features of the propagation of nonlinear acoustic waves propagating in
rotating dusty plasma will be modified. \ Understanding magnetized dusty
plasma has been a necessity and a challenge for both theoreticians and
experimentalists as for the former all charge-dependent forces and
fluctuating nature of dust charge can potentially be modified and latter for
the complexities involved. {\LARGE \ }In this regard many attempts have been
made, for instance K\"{a}hlert et al. (2012) proposed frictional coupling
between a dusty plasma and the neutral gas to mimic the dust magnetization
in a complex plasma. In this approach properties of the light species
electrons and ions were not affected; however, angular momentum from a
rotating gas column was transferred to a well-controlled rotation of the
dust cloud. In this way the induced Coriolis force $2m(\overrightarrow{v}%
\times \overrightarrow{\Omega })$ acting on objects moving with velocity $(%
\overrightarrow{v})$ when viewed in a rotating reference frame (with
frequency $\Omega )$ acts in a similar manner as the Lorentz force in a
magnetic field $Q(\overrightarrow{v}\times \overrightarrow{B})$ does \cite%
{26}. Needless to mention, the approach used by K\"{a}hlert et al. is
limited to study some particular phenomena and processes like dust charging,
formation of wakefield, Coulomb shielding, modulation and filamentation,
etc. cannot be studied with this. In an other attempt to study longitudinal
spectrum of collective excitations for different rotation rates with a \
high value of magnetic induction (\symbol{126}3200 T), authors spotted the
onset of the magnetoplasmon-like mode in a 2D single-layer dusty plasma \cite%
{27}. Both the studies in Ref. [26 \& 27] reported lower Coulomb coupling in
the rotating system leading to a liquid state compared to the nonrotating
case. Magnetorotational instability (MRI) also known as
Velikhov-Chandrasekhar instability which is a feature of purely rotating
fluids was investigated in differentially rotating dusty plasma where dust
particles were considered fixed \cite{28}. Conclusively magnetized and
rotating dusty plasmas are crucial. \ It is worth mentioning to mimic
laboratory and astrophysical settings group at Maryland is establishing
experiments for the high-velocity rotating dusty plasmas, in particular to
understand velocity limits and stability of dusty plasmas, and their
relation to high-temperature magnetized plasmas\cite{29}.

In the past, many times basic nonlinear properties of dust acoustic waves
propagating in the Maxwellian dusty plasma were studied. \ However data from
satellite observations predict the presence of super energetic long tails (
or shoulders at low energy) of nonthermal \ plasma particles \cite{30}, that
is why for the more accurate study of different stable/unstable collective
modes, nonthermal/non-Maxwellian distribution functions like Kappa \cite{31}%
, Cairns\cite{32} and generalized Lorentzian ($r,q)$ \cite{33} like
distributions have been used. Where features of nonthermality are contained
in the parameters (given as spectral indices) like $\kappa ,$ $\alpha ,$ $r$
and $q$ for the Kappa, Cairns and $r,q$ distributions, respectively. or
suprathermal tail.

In the present investigation, analytical model for obliquely propagating
nonlinear dust acoustic wave in a rotating magnetized dusty plasma will be
developed where electrons and ions will be treated as Kappa and Cairns
distributed For the plasma particles in thermal equilibrium, Kappa
distribution function is given as:

{\normalsize 
\begin{equation}
f_{s}^{\kappa }(v_{s})=\frac{n_{so}}{(2\pi )^{3/2}(\kappa -3/2)v_{ts}^{3}}%
\frac{\Gamma (1+\kappa )}{\Gamma (\kappa -1/2)}\left( 1+\frac{v_{s}^{2}}{%
2(\kappa -3/2)v_{ts}^{2})}\right) ^{-(1+\kappa )}  \tag{1}
\end{equation}%
}where $n_{so}$ represents the number density of the plasma species $%
s(=e,i,d $ for electron, ions and dust, respectively). The $%
v_{ts}=(T_{s}/m_{s})^{1/2} $ is the thermal speed, in which $T_{s}$ and $%
m_{s}$ denote the temperature in energy unit and mass, respectively. Where $%
\kappa $ measures the superthermality and $\Gamma $ is the well-known gamma
function. Condition of ${\normalsize \kappa }>3/2$ must hold in order to
have the realistic thermal speeds of plasma particles. Note that for the
larger values of kappa, Maxwellian distribution is retrieved. The Cairns
distribution function which was originally introduced after in situ
observations of Viking spacecraft and Freja satellite missions study
ion-sound cavitons like structures is given as: 
\begin{equation}
f_{s}^{C}(v_{s})=\frac{n_{so}}{(2\pi )^{3/2}(1+\alpha )v_{ts}^{3}}\left(
1+\alpha \frac{v_{s}^{4}}{v_{ts}^{4}}\right) exp\left( -\frac{v_{s}^{2}}{%
2v_{ts}^{2}}\right) ,  \tag{2}
\end{equation}%
where $\alpha $ determines the population of nonthermal plasma particles
i.e., for $\alpha \rightarrow 0$, one can achieve Maxwellian distribution.
of nonthermal energetic particles.

In this article, we will address dust acoustic shock wave which are useful
for the dusty plasma experiments under microgravity research and also
important from perspectives for future studies. We employ the well known
reductive perturbation technique to derive the nonlinear Zakharov--Kuznetsov
(ZK) \ equation for these waves at both fast and slow time scales. Both
electrons and ions are considered to follow kappa/Cairns distribution. The
solitary wave solution of this equation is obtained in Sec. IV. In Sec. V,
the results are presented and discussed, and finally, the conclusion is
presented in Sec. VI.

The paper is organized in the following manner: In Sec. II, the physical
assumption and description of the problem to be addressed is given. Sections
III and IV deal with the study of dust modified acoustic shock waves at fast
time scale and derivation of the Zakharov-Kuzensov equation for the dust
acoustic shock wave at slow time, respectively. Quantitative analysis is
provided in Section V, finally, Sec. VI describes the conclusions.

\section{Physical assumptions and description of the model}

We will make the following assumptions to formulate the physical problem:

\begin{enumerate}
\item The plasma under consideration is magnetized, homogeneous and
unbounded. The plasma constituents are electrons ($n_{e})$, ions ($n_{i})$,
and negatively charged dust grains ($n_{d}),$ and no collisions have been
taken into account between the particles. Charge on dust grains is negative.
The quasineutrality condition is given by 
\begin{equation}
n_{i}=n_{e}+Z_{d}n_{d}  \tag{3}
\end{equation}%
here $Z_{d}$ is the charge of dust grain, and $n_{s}=n_{s0}+\delta
n_{s}+\delta n_{s}^{L}$, while $n_{s0}$ represents {the} equilibrium
density. Superscript `$L$' refers to the ultra-low frequency for {the} dust
acoustic wave (DAW) in comparison with {the} higher frequency dust modified
acoustic wave (DMAW), and $\delta n_{s}$ gives the density perturbation on
the DAW time scale. {The parameter} $s$ denotes the species dust, ions or
electrons.

\item While treating this problem, we shall first consider the regime where
dust is in background and it is also the fast time process which will be
followed by slow time (DAW) dynamics. In the former regime,the dust mass is
ignored while in the latter, the dust species is activated.

\item The magnetic field is taken along z-axiz $B=B_{0}\widehat{z}$. When
dealing the slow time process, the wavelength $\lambda =2\pi /k$ is assumed
to be much smaller than the gyroradius of plasma particle which allows us to
take dust as magnetized and but electrons and ions are then treated
un-magnetized.

\item Also we know when $B$ is weak, the electron gyroradius is much smaller
than the size of grains and therefore variations in the dust charge is too
small. In this case, electrons will approach dust grain surface quite
rapidly along the direction of $B$ and therefore fast electrons responsible
for charging the grains may be treated as Boltzmannian. Thus for low
frequency dynamics in a magnetized plasma when $\omega /k\ll v_{ts}$ the
lighter species can obey Boltzmann distribution.

\item As stated in the introduction, we aim to adopt kappa and Cairns
distribution for the lighter species. The normalized number density for
kappa and Cairns distributed particles respectively is given by 
\begin{equation}
n_{e(i)}=\left[ 1\mp \left( \kappa -\frac{3}{2}\right) ^{-1}\frac{\phi }{%
\sigma }\right] ^{-\kappa +1/2}  \tag{4}
\end{equation}%
and%
\begin{equation}
n_{e(i)}=\left[ 1\mp \frac{\beta }{\sigma }\phi \pm \frac{\beta }{\sigma ^{2}%
}\phi ^{2}\right] e^{\pm \phi }  \tag{5}
\end{equation}%
where $\kappa $ is the spectral index measuring the deviation form
Maxwellian distribution, $\sigma =T_{i}/T_{e}$ and $\beta =4\alpha /\left(
1+3\alpha \right) .$ For the electrons in (5), $\sigma =1.$ For $\kappa
\rightarrow \infty (\alpha =0)$ Maxwellian distribution is achieved. It is
to be noted that (4) and (5) will be used only when dust is active in slow
time scale howeevr the first case when dust in in background (fast time
phenomena) only ions dynamics play the role only Eq. (4) will be sued. 

\item The time for the excitation of dust acoustic shock wave which is
nonlinear processes is much smaller than required for further substantial
variations in dust charge, and so dust charge can be taken constant.

\item As described in the introduction, we are considering rotating plasmas,
it is worth mentioning here that we will only (in case of slow time scale)
consider dust particles to be rotating whereas rotation of other
constituents electrons and ions is not important for us. Also it is
considered that both rotational and magnetic axis are misaligned.
\end{enumerate}

\section{Fast time scale phenomenon}

Here we consider excitation of dust modified shock wave a fast time process,
since, dust particles are extremely massive compared to the other
constituents so they stay in the background (steady and immovable), however,
their existence can be viewed via quasineutrality condition only. In the
fluid for ions which are dynamic here, we will incorporate ion bulk
viscosity which can only be ignored for the incompressible fluids. However,
for the acoustic shock waves, the plasma compressibility is essential and
therefore fluid equations \ are written as 
\begin{equation}
\frac{\partial n_{i}}{\partial t}+\mathbf{\nabla }\cdot \left( n_{i}\mathbf{v%
}_{i}\right) =0  \tag{6}
\end{equation}%
\begin{equation}
\left( \frac{\partial }{\partial t}+\mathbf{v}_{i}\cdot \mathbf{\mathbf{%
\nabla }}\right) \mathbf{v}_{i}=-\frac{1}{Z_{d}\delta ^{-1}}\mathbf{\nabla }%
\phi +\omega _{ci}\left( \mathbf{v_{i}\times \hat{z}}\right) +-\frac{\sigma 
}{Z_{d}\delta ^{-1}}\frac{\mathbf{\nabla }n\mathbf{_{i}}}{n_{i}}+\eta
_{i}\nabla ^{2}\mathbf{v_{i}}+\left( \eta _{i}+\mu _{i}\right) \mathbf{%
\nabla }\left( \mathbf{\nabla \cdot v}_{i}\right)  \tag{7}
\end{equation}%
\begin{equation}
\nabla ^{2}\phi =\delta ^{-1}\left( \mu n_{e1}-\delta n_{i1}\right)  \tag{8}
\end{equation}%
were in Eq. (7), $\eta _{i}$ and $%
{\mu}%
_{i}$ represent the kinematic and bulk viscosity for ions (also called the
second coefficient viscosity), respectively. Whereas $\mu
=n_{eo}/Z_{d}n_{do},$ $=m_{d}/m_{i},$ $\sigma =T_{i}/T_{e}$ and $\delta
=n_{io}/Z_{d}n_{do}.$According to the convenience of the problem being
addressed, for the normalization we use dust parameters. $\omega
_{ci}=eB_{0}/m_{i}c$ has been normalized by the $\omega _{pd}=\left( 4\pi
n_{d0}e^{2}Z_{d}^{2}/m_{d}\right) ^{1/2}$. The space and time variables have
been normalized by the $\lambda _{d}=\left( T_{e}\delta ^{-1}/4\pi
n_{d0}e^{2}Z_{d}\right) ^{1/2}$and $\omega _{pd}^{-1},$ respectively.
Moreover, $\eta _{i}$ and $\mu _{i}$ are normalized by $\omega _{pd}\lambda
_{d}^{2}$. Equations (6)-(8) in the Cartesian components for 3D nonlinear
DMSW can be written as 
\begin{equation}
\frac{\partial n_{i}}{\partial t}+\frac{\partial \left( n_{i}v_{ix}\right) }{%
\partial x}+\frac{\partial \left( n_{i}v_{iy}\right) }{\partial y}+\frac{%
\partial \left( n_{i}v_{iz}\right) }{\partial z}=0  \tag{9}
\end{equation}%
\begin{align}
\frac{\partial v_{ix}}{\partial t}+v_{ix}\frac{\partial v_{ix}}{\partial x}%
+v_{iy}\frac{\partial v_{ix}}{\partial y}+v_{iz}\frac{\partial v_{ix}}{%
\partial z}& =-\frac{1}{Z_{d}\delta ^{-1}}\frac{\partial \phi }{\partial x}%
+\omega _{ci}v_{iy}-\frac{\sigma }{Z_{d}\delta ^{-1}n_{i}}\frac{\partial
n_{i}}{\partial x}+\eta _{i}\left( \frac{\partial ^{2}}{\partial x^{2}}+%
\frac{\partial ^{2}}{\partial y^{2}}+\frac{\partial ^{2}}{\partial z^{2}}%
\right) v_{ix}  \notag \\
& +\left( \eta _{i}+\mu _{i}\right) \left( \frac{\partial ^{2}v_{ix}}{%
\partial x^{2}}+\frac{\partial ^{2}v_{iy}}{\partial x\partial y}+\frac{%
\partial ^{2}v_{iz}}{\partial x\partial z}\right)  \tag{10}
\end{align}%
\begin{align}
\frac{\partial v_{iy}}{\partial t}+v_{ix}\frac{\partial v_{iy}}{\partial x}%
+v_{iy}\frac{\partial v_{iy}}{\partial y}+v_{iz}\frac{\partial v_{iy}}{%
\partial z}& =-\frac{1}{Z_{d}\delta ^{-1}}\frac{\partial \phi }{\partial y}%
-\omega _{ci}v_{ix}-\frac{\sigma }{Z_{d}\delta ^{-1}n_{i}}\frac{\partial
n_{i}}{\partial y}+\eta _{i}\left( \frac{\partial ^{2}}{\partial x^{2}}+%
\frac{\partial ^{2}}{\partial y^{2}}+\frac{\partial ^{2}}{\partial z^{2}}%
\right) v_{iy}  \notag \\
& +\left( \eta _{i}+\mu _{i}\right) \left( \frac{\partial ^{2}v_{ix}}{%
\partial y\partial x}+\frac{\partial ^{2}v_{iy}}{\partial y^{2}}+\frac{%
\partial ^{2}v_{iz}}{\partial y\partial z}\right)  \tag{11}
\end{align}%
\begin{align}
\frac{\partial \left( v_{iz}\right) }{\partial t}+v_{ix}\frac{\partial
\left( v_{iz}\right) }{\partial x}+v_{iy}\frac{\partial \left( v_{iz}\right) 
}{\partial y}+v_{iz}\frac{\partial \left( v_{iz}\right) }{\partial z}& =-%
\frac{1}{Z_{d}\delta ^{-1}}\frac{\partial \phi }{\partial z}-\frac{\sigma }{%
Z_{d}\delta ^{-1}n_{i}}\frac{\partial n_{i}}{\partial z}+\eta _{i}\left( 
\frac{\partial ^{2}}{\partial x^{2}}+\frac{\partial ^{2}}{\partial y^{2}}+%
\frac{\partial ^{2}}{\partial z^{2}}\right) v_{iz}  \notag \\
& +\left( \eta _{i}+\mu _{i}\right) \left( \frac{\partial ^{2}v_{ix}}{%
\partial z\partial x}+\frac{\partial ^{2}v_{iy}}{\partial z\partial y}+\frac{%
\partial ^{2}v_{iz}}{\partial z^{2}}\right)  \tag{12}
\end{align}%
\begin{equation}
\frac{\partial ^{2}\phi }{\partial x^{2}}+\frac{\partial ^{2}\phi }{\partial
y^{2}}+\frac{\partial ^{2}\phi }{\partial z^{2}}=\delta ^{-1}\left( \mu
+c_{1}\phi +c_{2}\phi ^{2}-\delta n_{i1}\right)  \tag{13}
\end{equation}%
To study small and finite amplitude DMSW we will use reductive perturbation
method \cite{15} and introduce stretched coordinates given as: 
\begin{align}
\xi & =\epsilon ^{1/2}x,\text{\ \ \ \ \ \ \ \ \ }\eta =\epsilon ^{1/2}y, 
\notag \\
\zeta & =\epsilon ^{1/2}\left( z-\lambda _{0}t\right) \text{ and }\tau
=\epsilon ^{3/2}t  \tag{15}
\end{align}%
where $\lambda _{0}$ is the speed with which shock wave propagates and a
dimensionless parameter, $\epsilon $ $\left( 0<\epsilon \ll 1\right) $
measures the strength of nonlinearity. Further, the other variables like
density, velocity, potential are expressed as: 
\begin{align}
n_{i}& =1+\varepsilon n_{1}+\epsilon ^{2}n_{2}+...,  \notag \\
v_{ix}& =\epsilon ^{3/2}u_{1}+\epsilon ^{2}u_{2}+...,  \notag \\
v_{iy}& =\epsilon ^{3/2}v_{1}+\epsilon ^{2}v_{2}+...,  \notag \\
v_{iz}& =v_{0}+\epsilon w_{1}+\epsilon ^{2}w_{2}+...,  \notag \\
\phi & =\epsilon \phi _{1}+\epsilon ^{2}\phi _{2}+...,  \tag{16}
\end{align}%
The ion kinematic viscosity, for weakly damped systems, is assumed to be
small, and can be expressed as 
\begin{equation}
\eta _{i}=\epsilon ^{1/2}\eta _{0},\text{ \ \ }\mu _{i}=\epsilon ^{1/2}\mu
_{0},  \tag{17}
\end{equation}%
where $\eta _{0}$ and $\mu _{0}$ have dimensions of a unit. Substitution of
equations (15)-(17) into (9)-(13) and collection of lowest order terms such
as $\epsilon ^{1}$ and $\epsilon ^{3/2}$ result in 
\begin{equation}
w_{1}=n_{1}\left( \lambda _{o}-v_{o}\right)  \tag{18}
\end{equation}%
\begin{equation}
n_{1}=\frac{c_{1}}{\delta }\phi _{1}  \tag{19}
\end{equation}%
\begin{equation}
\frac{1}{Z_{d}\delta ^{-1}}\frac{\partial \phi _{1}}{\partial \xi }+\frac{%
\sigma }{Z_{d}\delta ^{-1}}\frac{\partial n_{1}}{\partial \xi }-\omega
_{ci}v_{i}=0  \tag{20}
\end{equation}%
\begin{equation}
\frac{1}{Z_{d}\delta ^{-1}}\frac{\partial \phi _{1}}{\partial \eta }+\frac{%
\sigma }{Z_{d}\delta ^{-1}}\frac{\partial n_{1}}{\partial \eta }+\omega
_{ci}u_{i}=0  \tag{21}
\end{equation}%
\begin{equation}
-\left( \lambda _{o}-v_{o}\right) \frac{\partial w_{1}}{\partial \zeta }+%
\frac{\sigma }{Z_{d}\delta ^{-1}}\frac{\partial n_{1}}{\partial \zeta }+%
\frac{1}{Z_{d}\delta ^{-1}}\frac{\partial \phi _{1}}{\partial \zeta }=0 
\tag{22}
\end{equation}%
From above equations, we obtain linear dispersion relation for the dust
modified ion acoustic wave%
\begin{equation}
\lambda _{0}=v_{0}\pm \sqrt{\frac{1}{Z_{d}\delta ^{-1}c_{1}}\left( \sigma
c_{1}+\delta \right) }=v_{0}\pm \left[ \frac{n_{i0}}{Z_{d}^{2}n_{d0}}\frac{%
T_{i}}{T_{e}}\left( 1+\frac{T_{e}}{c_{1}Z_{d}T_{i}}\right) \right] ^{1/2} 
\tag{23}
\end{equation}%
above equation describes the phase speed of the waves, where $+(-)$ sign
refers to the fast (slow) modes. It can be observed from (23) that linear
phase velocity of the DMSW is not affected by the magnetic field and
viscosity; however, presence of the factor $n_{i0}/Z_{d}n_{d0}$ shows that
when dust is present in the background the phase velocity of the fast
ion-acoustic mode increases whereas it is reduced for the slow mode.

Collection of higher order terms such as $\epsilon ^{2}$ and $\epsilon
^{5/2} $ return us 
\begin{equation}
\left( \lambda _{0}-v_{0}\right) \frac{\partial u_{1}}{\partial \zeta }%
+\omega _{ci}v_{2}=0  \tag{24}
\end{equation}%
\begin{equation}
\left( \lambda _{0}-v_{0}\right) \frac{\partial v_{1}}{\partial \zeta }%
=\omega _{ci}u_{2}  \tag{25}
\end{equation}%
\begin{equation}
\frac{\partial ^{2}\phi _{1}}{\partial \xi ^{2}}+\frac{\partial ^{2}\phi _{1}%
}{\partial \eta ^{2}}+\frac{\partial ^{2}\phi _{1}}{\partial \zeta ^{2}}%
=\delta ^{-1}\left( c_{1}\phi _{2}+c_{2}\phi _{1}^{2}-\delta n_{2}\right) 
\tag{26}
\end{equation}%
\begin{equation}
-\left( \lambda _{0}-v_{0}\right) \frac{\partial n_{2}}{\partial \zeta }+%
\frac{\partial u_{2}}{\partial \xi }+\frac{\partial v_{2}}{\partial \eta }+%
\frac{\partial w_{2}}{\partial \zeta }=-\frac{\partial n_{1}}{\partial \tau }%
-\frac{\partial }{\partial \zeta }\left( n_{1}w_{1}\right)  \tag{27}
\end{equation}%
\begin{multline}
-\left( \lambda _{0}-v_{0}\right) \frac{\partial w_{2}}{\partial \zeta }+%
\frac{\partial w_{1}}{\partial \tau }+w_{1}\frac{\partial w_{1}}{\partial
\zeta }+\frac{1}{Z_{d}\delta ^{-1}}\frac{\partial \phi _{2}}{\partial \zeta }%
+\frac{\sigma }{Z_{d}\delta ^{-1}}\frac{\partial n_{2}}{\partial \zeta }=%
\frac{\sigma }{Z_{d}\delta ^{-1}}n_{1}\frac{\partial n_{1}}{\partial \zeta }
\notag \\
+\eta _{0}\left( \frac{\partial ^{2}}{\partial \xi ^{2}}+\frac{\partial ^{2}%
}{\partial \eta ^{2}}+\frac{\partial ^{2}}{\partial \zeta ^{2}}\right)
w_{1}+\left( \eta _{0}+\mu _{0}\right) \frac{\partial ^{2}w_{1}}{\partial
\zeta ^{2}}  \tag{28}
\end{multline}%
From Eqs. (24-28), we obtain the ZKB equation describing the dust modified
ion acoustic shock wave 
\begin{equation}
\frac{\partial \phi _{1}}{\partial \tau }+A\phi _{1}\frac{\partial \phi _{1}%
}{\partial \zeta }+B\frac{\partial ^{3}\phi _{1}}{\partial \zeta ^{3}}+C%
\frac{\partial }{\partial \zeta }\left( \frac{\partial ^{2}\phi _{1}}{%
\partial \xi ^{2}}+\frac{\partial ^{2}\phi _{1}}{\partial \eta ^{2}}\right)
-D\left( \frac{\partial ^{2}\phi _{1}}{\partial \xi ^{2}}+\frac{\partial
^{2}\phi _{1}}{\partial \eta ^{2}}+\frac{\partial ^{2}\phi _{1}}{\partial
\zeta ^{2}}\right) -E\frac{\partial ^{2}\phi _{1}}{\partial \zeta ^{2}}=0 
\tag{29}
\end{equation}%
where%
\begin{equation}
A=\frac{c_{1}}{\delta }\left[ \frac{1}{2}+(\lambda _{o}-v_{o})+\frac{%
c_{2}\delta ^{3}}{Z_{d}c_{1}^{3}(\lambda _{o}-v_{o})}-\frac{T_{i}}{%
2Z_{d}T_{e}(\lambda _{o}-v_{o})}\right]  \tag{30}
\end{equation}%
\begin{equation}
B=\frac{\delta ^{3}}{2Z_{d}(\lambda _{o}-v_{o})c_{1}^{2}}  \tag{31}
\end{equation}%
\begin{equation}
C=\frac{\delta T_{i}(\lambda _{o}-v_{o})}{2\omega _{ci}^{2}Z_{d}T_{e}}+\frac{%
\delta ^{2}}{2c_{1}\omega _{ci}^{2}Z_{d}}+\frac{\delta ^{3}}{%
2c_{1}^{2}(\lambda _{o}-v_{o})Z_{d}}  \tag{32}
\end{equation}%
\begin{equation}
D=\frac{\eta _{0}}{2},\text{ }E=\frac{\eta _{0}+\mu _{0}}{2}  \tag{33}
\end{equation}%
where $A$ is the nonlinear coefficient, $B$ and $C$ \ are dispersive whereas 
$D$ and $E$ represent dissipation coefficients.

Presence of the factor $n_{i0}/Z_{d}n_{d0}$ in (30) shows that nonlinear
coefficient significantly effeced by the dust in background. The reducing
4th term is small as $\delta $ is in the denominator and the factor $%
T_{i}/T_{e}$ is also less than unity in usual astrophysical and lab
enviornments whereas it can only be larger than unity in tokamaks.It is
obvious from (33) that dissipation coefficients are not affected by the
presence of dust in the background.

To examine the shock like solution of ZKB equation (29), we introduce the
parameter $\chi $ as%
\begin{equation}
\chi =l_{x}\xi +l_{y}\eta +l_{z}\zeta -U_{0}\tau  \tag{34}
\end{equation}%
here $l_{\alpha =x,y,z}$ are the direction cosines and $U_{0}$ is wave speed
for the nonlinear propagation. Using Eq. (34) into (29) yields the following
ordinary differential equation (ODE) as 
\begin{equation}
-U_{0}\frac{d\phi _{1}}{d\chi }+Al_{z}\phi _{1}\frac{d\phi _{1}}{d\chi }%
+Hl_{z}\frac{d^{3}\phi _{1}}{d\chi ^{3}}-G\frac{d^{2}\phi _{1}}{d\chi ^{2}}%
=0,  \tag{35}
\end{equation}%
where $H=l_{z}^{2}B+\left( l_{x}^{2}+l_{y}^{2}\right) C$ and $G=El_{z}^{2}+D$%
. The shock like solution of Eq. (35) can be found using hyperbolic tangent
method \cite{34}\textbf{. }Thus, employing the condition that $\phi _{1}$ is
bounded at $\chi =\pm \infty $, we obtain shock wave solution%
\begin{equation}
\phi _{1}(\chi )=\frac{3}{25}\text{ }\frac{G^{2}}{HAl_{z}^{2}}\left[ 2-2%
\text{ tanh}\left( \frac{G}{10Hl_{z}}\chi \right) +\text{sech}^{2}\left( 
\frac{G}{10Hl_{z}}\chi \right) \right]  \tag{36}
\end{equation}%
As is evident from the above equation shock \ is formed due to the ion
kinematic viscosity term. Here, $10Hl_{z}/G$ and $\left( 9/25\right) \left(
G^{2}/HAl_{z}^{2}\right) $ represent the width and amplitude (depends upon A
the nonlinear coefficient) of the shock structure, respectively.

\section{Dust acoustic (DA) wave at slow time scale}

In this section we derive dispersion (linear and nonlinear) of the dust
acoustic waves and take into account the quasi-neutrality condition $\delta
n_{i}\sim \delta n_{e}+Z_{d}\delta n_{d}$, since the time with which
velocity and density of lighter species vary is much shorter than that of
heavier dust i.e., 
\begin{equation}
t_{i}\left( \sim \frac{1}{\omega _{pi}}\right) \sim v_{e}\left( \frac{%
\partial v_{e}}{\partial t}\right) ^{-1},n_{e}\left( \frac{\partial n_{e}}{%
\partial t}\right) ^{-1}<<t_{d}\left( \sim \frac{1}{\omega _{pd}}\right) 
\tag{37}
\end{equation}%
In this case, the dynamic effects of {the} dust grains are included, because
we are interested in the dust's time and space scales, so the fluid
equations for the dust are 
\begin{equation}
\frac{\partial n_{d}}{\partial t}+\mathbf{\nabla }\cdot \left( n_{d}\mathbf{v%
}_{d}\right) =0  \tag{38}
\end{equation}%
\begin{equation}
\left( \frac{\partial }{\partial t}+\mathbf{v}_{d}\cdot \mathbf{\mathbf{%
\nabla }}\right) \mathbf{v}_{d}=\delta \mathbf{\nabla }\phi -\omega
_{cd}\left( \mathbf{v_{d}\times \hat{z}}\right) -\frac{\sigma _{d}}{\delta
^{-1}}\frac{\mathbf{\nabla n_{d}}}{n_{d}}+2\Omega _{o}\left( \mathbf{%
v_{d}\times \hat{z}}\right) +\eta _{d}\nabla ^{2}\mathbf{v_{d}}+\left( \eta
_{d}+\mu _{d}\right) \mathbf{\nabla }\left( \mathbf{\nabla \cdot v}%
_{d}\right)  \tag{39}
\end{equation}%
\begin{equation}
\nabla ^{2}\phi =\delta ^{-1}\left( \mu n_{e}-\delta n_{i}+n_{d}\right) 
\tag{40}
\end{equation}%
where $\sigma _{d}=T_{d}/T_{e}Z_{d},$ $\Omega _{o}$ is the rotational
frequency of the dust. Dust fluid velocity $\mathbf{v}_{d}$ is normalized by
dust acoustic speed $c_{s}=\left( Z_{d}T_{e}\delta ^{-1}/m_{d}\right) ^{1/2}$%
. the electrostatic potential $\phi $ is normalized by $T_{e}/e$. Also, the
dust kinematic viscosity $\eta _{d}$ and the second coefficient of viscosity 
$\mu _{d}$ are normalized by $\omega _{pd}\lambda _{d}^{2}$. The ions and
electrons are considered to obey Kappa and Cairns distributions, their
number densities given by Eqs. (4) and (5). Above equations can be expressed
in the Cartesian components form as follows%
\begin{equation}
\frac{\partial n_{d}}{\partial t}+\frac{\partial \left( n_{d}v_{dx}\right) }{%
\partial x}+\frac{\partial \left( n_{d}v_{dy}\right) }{\partial y}+\frac{%
\partial \left( n_{d}v_{dz}\right) }{\partial z}=0  \tag{41}
\end{equation}%
\begin{align}
\frac{\partial v_{dx}}{\partial t}+v_{dx}\frac{\partial v_{dx}}{\partial x}%
+v_{dy}\frac{\partial v_{dx}}{\partial y}+v_{dz}\frac{\partial v_{dx}}{%
\partial z}& =\delta \frac{\partial \phi }{\partial x}-\Omega _{c}v_{dy}-%
\frac{\sigma _{d}\delta }{n_{d}}\frac{\partial n_{d}}{\partial x}+\eta
_{d}\left( \frac{\partial ^{2}}{\partial x^{2}}+\frac{\partial ^{2}}{%
\partial y^{2}}+\frac{\partial ^{2}}{\partial z^{2}}\right) v_{dx}  \notag \\
& +\left( \eta _{d}+\mu _{d}\right) \left( \frac{\partial ^{2}v_{dx}}{%
\partial x^{2}}+\frac{\partial ^{2}v_{dy}}{\partial x\partial y}+\frac{%
\partial ^{2}v_{dz}}{\partial x\partial z}\right)  \tag{42}
\end{align}%
\begin{align}
\frac{\partial v_{dy}}{\partial t}+v_{dx}\frac{\partial v_{dy}}{\partial x}%
+v_{dy}\frac{\partial v_{dy}}{\partial y}+v_{dz}\frac{\partial v_{dy}}{%
\partial z}& =\delta \frac{\partial \phi }{\partial y}+\Omega _{c}v_{dx}-%
\frac{\sigma _{d}\delta }{n_{d}}\frac{\partial n_{d}}{\partial y}+\eta
_{d}\left( \frac{\partial ^{2}}{\partial x^{2}}+\frac{\partial ^{2}}{%
\partial y^{2}}+\frac{\partial ^{2}}{\partial z^{2}}\right) v_{dy}  \notag \\
& +\left( \eta _{d}+\mu _{d}\right) \left( \frac{\partial ^{2}v_{dx}}{%
\partial y\partial x}+\frac{\partial ^{2}v_{dy}}{\partial y^{2}}+\frac{%
\partial ^{2}v_{dz}}{\partial y\partial z}\right)  \tag{43}
\end{align}%
\begin{align}
\frac{\partial \left( v_{dz}\right) }{\partial t}+v_{dx}\frac{\partial
\left( v_{dz}\right) }{\partial x}+v_{dy}\frac{\partial \left( v_{dz}\right) 
}{\partial y}+v_{dz}\frac{\partial \left( v_{dz}\right) }{\partial z}&
=\delta \frac{\partial \phi }{\partial z}-\frac{\sigma _{d}\delta }{n_{d}}%
\frac{\partial n_{d}}{\partial z}+\eta _{d}\left( \frac{\partial ^{2}}{%
\partial x^{2}}+\frac{\partial ^{2}}{\partial y^{2}}+\frac{\partial ^{2}}{%
\partial z^{2}}\right) v_{dz}  \notag \\
& +\left( \eta _{d}+\mu _{d}\right) \left( \frac{\partial ^{2}v_{dx}}{%
\partial z\partial x}+\frac{\partial ^{2}v_{dy}}{\partial z\partial y}+\frac{%
\partial ^{2}v_{dz}}{\partial z^{2}}\right)  \tag{44}
\end{align}%
\begin{equation}
\frac{\partial ^{2}\phi }{\partial x^{2}}+\frac{\partial ^{2}\phi }{\partial
y^{2}}+\frac{\partial ^{2}\phi }{\partial z^{2}}=\delta ^{-1}\left(
+c_{d1}\phi +c_{d2}\phi ^{2}-1+n_{d}\right)  \tag{45}
\end{equation}%
where we define $\Omega _{c}=\omega _{cd}-2\Omega _{o}$ and $\mu -\delta =1$%
, comes from the charge neutrality condition and 
\begin{align}
c_{d1}& =\left\{ 
\begin{array}{c}
\left( \mu -\frac{\delta }{\sigma }\right) \left( 1+\beta \right) \text{ }(%
\text{Cairns}), \\ 
\frac{\mu \left( \kappa -1/2\right) }{\kappa -3/2}+\frac{\delta \left(
\kappa -1/2\right) }{\left( \kappa -3/2\right) \sigma }\text{ }(\text{kappa})%
\end{array}%
\right.  \notag \\
c_{d2}& =\left\{ 
\begin{array}{c}
\frac{\mu }{2}-\frac{\delta (1+4\beta )}{2\sigma ^{2}}\text{ }(\text{Cairns}%
), \\ 
\left( \mu -\frac{\delta }{\sigma ^{2}}\right) \frac{\mu \left( \kappa
-1/2\right) \left( \kappa +1/2\right) }{2\left( \kappa -3/2\right) ^{2}}%
\text{ }(\text{kappa})\text{ }%
\end{array}%
\right.  \tag{46}
\end{align}%
Like in previous section, we opt standard reductive perturbation method and
introduce stretched coordinates to obtain the ZK-Burgers equation 
\begin{align}
\xi & =\epsilon ^{1/2}x,\text{\ \ \ \ \ \ \ \ \ }\eta =\epsilon ^{1/2}y, 
\notag \\
\zeta & =\epsilon ^{1/2}\left( z-\lambda _{0d}t\right) \text{ and }\tau
=\epsilon ^{3/2}t  \tag{47}
\end{align}%
where $\lambda _{0d}$ is the propagation speed of the dust acoustic wave to
be determined later. Furthermore, the dependent variables $n_{d}$, $\mathbf{v%
}_{d},$and $\phi $ are expanded in power series of $\epsilon $ as%
\begin{align}
n_{d}& =1+\varepsilon n_{d1}+\epsilon ^{2}n_{d2}+...,  \notag \\
v_{dx}& =\epsilon ^{3/2}u_{d1}+\epsilon ^{2}u_{d2}+...,  \notag \\
v_{dy}& =\epsilon ^{3/2}v_{d1}+\epsilon ^{2}v_{d2}+...,  \notag \\
v_{dz}& =v_{d0}+\epsilon w_{d1}+\epsilon ^{2}w_{d2}+...,  \notag \\
\phi & =\epsilon \phi _{1}+\epsilon ^{2}\phi _{2}+...,  \tag{48}
\end{align}%
Similarly we express 
\begin{align}
\eta _{d}& =\epsilon ^{1/2}\eta _{d0}  \notag \\
\mu _{d}& =\epsilon ^{1/2}\mu _{d0}  \tag{49}
\end{align}%
Collection of lowest order terms return us \ the following equations: 
\begin{equation}
w_{d1}=n_{d1}\left( \lambda _{do}-v_{do}\right)  \tag{50}
\end{equation}%
\begin{equation}
n_{d1}=-c_{d1}\phi _{1}  \tag{51}
\end{equation}%
\begin{equation}
\delta \frac{\partial \phi _{1}}{\partial \xi }-\sigma _{d}\delta \frac{%
\partial n_{d1}}{\partial \xi }-\Omega _{c}v_{d1}=0  \tag{52}
\end{equation}%
\begin{equation}
\delta \frac{\partial \phi _{1}}{\partial \eta }-\sigma _{d}\delta \frac{%
\partial n_{d1}}{\partial \eta }+\Omega _{c}u_{d1}=0  \tag{53}
\end{equation}%
\begin{equation}
-\left( \lambda _{do}-v_{do}\right) \frac{\partial w_{d1}}{\partial \zeta }%
-\sigma _{d}\delta \frac{\partial n_{d1}}{\partial \zeta }+\delta \frac{%
\partial \phi _{1}}{\partial \zeta }=0  \tag{54}
\end{equation}%
From equations (51-54), we obtain%
\begin{equation}
\lambda _{d0}=v_{d0}\pm \sqrt{\frac{\delta (1+\sigma _{d}c_{d1})}{c_{d1}}} 
\tag{55}
\end{equation}%
Equation (55) represents the phase velocity for dust acoustic waves and $\pm 
$ sign for the fast and slow dust acoustic speeds. Higher order term in $%
\epsilon $ gives us 
\begin{equation}
\left( \lambda _{d0}-v_{d0}\right) \frac{\partial u_{d1}}{\partial \zeta }%
-\Omega _{c}v_{d2}=0  \tag{56}
\end{equation}%
\begin{equation}
\left( \lambda _{d0}-v_{d0}\right) \frac{\partial v_{d1}}{\partial \zeta }%
+\Omega _{c}u_{d2}=0  \tag{57}
\end{equation}%
\begin{equation}
\frac{\partial ^{2}\phi _{1}}{\partial \xi ^{2}}+\frac{\partial ^{2}\phi _{1}%
}{\partial \eta ^{2}}+\frac{\partial ^{2}\phi _{1}}{\partial \zeta ^{2}}%
=\delta ^{-1}\left( c_{d1}\phi _{2}+c_{d2}\phi _{1}^{2}+n_{d2}\right) 
\tag{58}
\end{equation}%
\begin{equation}
-\left( \lambda _{d0}-v_{d0}\right) \frac{\partial n_{d2}}{\partial \zeta }+%
\frac{\partial u_{d2}}{\partial \xi }+\frac{\partial v_{d2}}{\partial \eta }+%
\frac{\partial w_{d2}}{\partial \zeta }=-\frac{\partial n_{d1}}{\partial
\tau }-\frac{\partial }{\partial \zeta }\left( n_{d1}w_{d1}\right)  \tag{59}
\end{equation}%
\begin{multline}
-\left( \lambda _{d0}-v_{d0}\right) \frac{\partial w_{d2}}{\partial \zeta }+%
\frac{\partial w_{d1}}{\partial \tau }+w_{d1}\frac{\partial w_{d1}}{\partial
\zeta }-\delta \frac{\partial \phi _{2}}{\partial \zeta }+\delta \sigma _{d}%
\frac{\partial n_{d2}}{\partial \zeta }=\delta \sigma _{d}n_{d1}\frac{%
\partial n_{d1}}{\partial \zeta }  \notag \\
+\eta _{d0}\left( \frac{\partial ^{2}}{\partial \xi ^{2}}+\frac{\partial ^{2}%
}{\partial \eta ^{2}}+\frac{\partial ^{2}}{\partial \zeta ^{2}}\right)
w_{1}+\left( \eta _{d0}+\mu _{d0}\right) \frac{\partial ^{2}w_{d1}}{\partial
\zeta ^{2}}  \tag{60}
\end{multline}%
Finally trivial algebra steps return us ZK-Burgers equation in the form 
\begin{equation}
\frac{\partial \phi _{1}}{\partial \tau }+A^{^{\prime }}\phi _{1}\frac{%
\partial \phi _{1}}{\partial \zeta }+B^{^{\prime }}\frac{\partial ^{3}\phi
_{1}}{\partial \zeta ^{3}}+C^{^{\prime }}\frac{\partial }{\partial \zeta }%
\left( \frac{\partial ^{2}\phi _{1}}{\partial \xi ^{2}}+\frac{\partial
^{2}\phi _{1}}{\partial \eta ^{2}}\right) -D^{^{\prime }}\left( \frac{%
\partial ^{2}\phi _{1}}{\partial \xi ^{2}}+\frac{\partial ^{2}\phi _{1}}{%
\partial \eta ^{2}}+\frac{\partial ^{2}\phi _{1}}{\partial \zeta ^{2}}%
\right) -E^{^{\prime }}\frac{\partial ^{2}\phi _{1}}{\partial \zeta ^{2}}=0 
\tag{61}
\end{equation}%
where 
\begin{equation}
A^{\prime }=\frac{\sigma _{d}c_{d1}-2c_{2}c_{d1}^{2}-3c_{d1}\alpha
_{d}(\lambda _{do}-v_{do})^{2}}{2\delta ^{-1}(\lambda _{do}-v_{do})} 
\tag{62}
\end{equation}%
\begin{equation}
B^{^{\prime }}=\frac{\delta ^{2}}{2c_{d1}^{2}(\lambda _{do}-v_{do})} 
\tag{63}
\end{equation}%
\begin{equation}
C^{^{\prime }}=\frac{1+\Omega _{c}^{-2}\delta ^{-1}c_{d1}(c_{d1}\sigma
_{d}+1)(\lambda _{do}-v_{do})^{2}}{2\delta ^{-2}c_{d1}^{2}(\lambda
_{do}-v_{do})}  \tag{64}
\end{equation}%
\begin{equation}
D^{^{\prime }}=\frac{\eta _{0}}{2},\text{ }E^{\prime }=\frac{\eta _{0}+\mu
_{0}}{2}  \tag{65}
\end{equation}%
Hence the Burger terms having coefficients ($D^{^{\prime }}$ and $%
E^{^{\prime }}$) which are responsible for the generation of shock wave,
originates due to the viscosity term. Also note that since dissipative terms
are$\ $proportional to $\eta _{0}$ and $\mu _{0},$it mean in the absence of
this the dissipative terms would vanish and ZK-Burger equation will be
reduced to usual ZK equation admitting nonlinear soliton solutions only.

\section{Stationary solution and quantitative analysis}

In this section, we numerically solve Eq. $\left( 61\right) $ to examine
dust acoustic shock waves for kappa and Cairns distributed ions and
electrons.

Shock like structures \ can be studied if the coefficients of the Burger
term arising from viscous nature of plasma are positive, i.e. ($D^{^{\prime
}},E^{\prime }>0$). To obtain the solution of ZKB equation (61), we first
transform (61) into another form using $\chi =\gamma _{x}\xi +\gamma
_{y}\eta +\gamma _{z}\zeta -U_{d}\tau $. Where $\gamma _{s=x,y,z}$ are the
direction cosines and $U_{d}$ is now normalized to $C_{sd}$. This yields: 
\begin{equation}
-U_{d}\frac{d\phi _{1}}{d\chi }+A^{^{\prime }}\gamma _{z}\phi _{1}\frac{%
d\phi _{1}}{d\chi }+H^{^{\prime }}\gamma _{z}\frac{d^{3}\phi _{1}}{d\chi ^{3}%
}-G^{^{\prime }}\frac{d^{2}\phi _{1}}{d\chi ^{2}}=0,  \tag{67}
\end{equation}%
where $H^{^{\prime }}=\gamma _{z}^{2}B^{^{\prime }}+\left( \gamma
_{x}^{2}+\gamma _{y}^{2}\right) C^{^{\prime }}$ and $G^{^{\prime
}}=E^{^{\prime }}\gamma _{z}^{2}+D^{^{\prime }}$. Again employing the
hyperbolic tangent (tanh) method along with the boundary conditions we found
the shock wave solutions 
\begin{equation}
\phi _{1}(\chi )=\frac{3}{25}\text{ }\frac{G^{^{\prime }2}}{H^{^{\prime
}}A^{^{\prime }}\gamma _{z}^{2}}\left[ 2-2\text{ tanh}\left( \frac{%
G^{^{\prime }}}{10H^{^{\prime }}\gamma _{z}}\chi \right) +\text{sech}%
^{2}\left( \frac{G^{^{\prime }}}{10H^{^{\prime }}\gamma _{z}}\chi \right) %
\right]  \tag{68}
\end{equation}%
with, $10H^{^{\prime }}\gamma _{z}G^{^{\prime }-1}$ providing the width and $%
\frac{9}{25}G^{2}H^{-1}A\gamma _{z}^{-2}$ gives the amplitude of shock waves
moving with speed $U_{d}$. The shock width and amplitude are dependent on
the dispersive coefficients $B^{^{\prime }}$ and $C^{^{\prime }}$,
dissipative coefficients $D^{^{\prime }}$ and $E^{^{\prime }}$, and
direction cosines $\gamma _{x},$ $\gamma _{y},$ $\gamma _{z}$. The amplitude
is also dependent on the nonlinear coefficient $A^{^{\prime }}.$ The
dependence of these coefficients on various plasma parameter determines the
shape of the shock profile.

\textbf{Linear dispersion relation }

Now we numerically solve the Eq. (61). For illustration we have chosen some
typical parameters of the dusty plasmas

$Z_{d}=50$, $n_{do}=1cm^{-3}$, $n_{io}=6\times 10^{3}cm^{-3},$ $l_{z}=0.9$, $%
T_{i}=2$eV, and $T_{e}=8$eV. First of all we study how the phase velocity ($%
\lambda _{do}$) of the dust acoustic wave will be modified for the
nonthermal distribution parameters $\kappa $ and $\alpha $ ( kappa/Cairns)
distributions.

It can be seen from Fig. (1) that phase velocity $increases$ in case of fast
mode upon increment in the value of $\kappa $ and vice versa for the slow
mode.

In Figs. (2) and (3) we observe that for the fixed flow speed $%
v_{0}=2.3\times 10^{2}cms^{-1},$ increasing the value of $\alpha $ from $%
0.1-1.7$ leads to enhancement in the phase velocity ($\lambda _{do})$ for
the fast as well as slow modes of dust acoustic wave at slow time scale.
Whereas in another graph\ [Fig. (4)]. we vary the flow speed for the fast
mode only such as $v_{0}=2\times 10^{2}cms^{-1}$, $v_{0}=2.3\times
10^{2}cms^{-1},$ $v_{0}=2.6\times 10^{2}cms^{-1}$ and note how this affects
the phase velocity of fast mode whereas $\alpha $ has been chosen between $0$
and $0.5.$

\textbf{Effect on Coefficients}

It is shown in Fig. (5) that on increasing $\alpha $ between $1-5$, the
nonlinear coefficient $A^{\prime }$ increases. In another figure (6), we
show how dispersive coefficient $B^{^{\prime }}$ changes with the variations
in the parameter $\sigma _{d}=T_{d}/T_{e}Z_{d}$, we note that incraesing
this factor $\sigma _{d}$ causes reduction in the strenth of the $%
B^{^{\prime }}$.

In Fig .(7), we observe how the nonlinear coefficient ($A^{\prime }$)
changes while changing the value of charge number that is $Z_{d}$. It is
depicted from the figure that incaresing the $Z_{d}$ from $20$ to $72$, the
nonlinear coefficient decreases whereas after 72 it rises abruptly.
Similarly for the values below 20, $A^{\prime }$ incraeses.

\textbf{Effect of viscosity}

In Fig. (8), we examine the behavior of shock structure for different values
of ion kinematic viscosity coefficient and observe that for fixed value of $%
\kappa (=3)$, upon increasing the values of viscosity $\eta
_{o}(=0.10,0.12,0.14)$, strength of the potential is enhanced significantly.
In this case the shcok wave formed is of compressive in nature. However we
also observe in Fig. (9) that enhancing the percentage of the suprathermal
electrons i.e. $\kappa $ $(=2,3,4)$ for this fixed viscosity ($\eta _{o}=.09$%
), the shock amplitude increases when there is a decrease in the the kappa.
This is interesting and show supper thermality affects shocks. 

Main conclusion from here is both enhancing superthermality and kinematic
viscosity both affect significantly shock waves.

In case of Cairns distribution Fig. (10) $\eta _{o}(=0.06,0.08,0.1)$ for
fixed $\alpha =0.15,$ $\Omega _{c}=0.3,$ we again observe a significant
change in the amplitude of the shock wave however in this case shock wave is
of rarefaction nature.

\textbf{Effect of rotation}

Analysis of the coefficients of the ZK-Burger equation shows that rotation
contributes only in the dispersive coefficient. \ Therefore rotational
frequency and the external magnetic field do not directly affect the
amplitude of the shock wave; however, they do on the width of the shock
waves. It can be seen  in Fig. (11), as we increase both the rotation $%
(\Omega _{c}=0.1,0.2,0.4,0.6)$ and keep viscosity and kappa fixed as $\eta
_{o}=0.10,$ $\kappa =3,$ a rarefaction shock structure will be formed. 

We also note in Fig. (12) that in case of Cairns distribution for $\eta
_{o}=0.1$ for $\alpha =0.45,$ and $\Omega _{c}=0.1,0.2,0.6,$we observe that
initially the perturbed potential $\phi _{1}$ increases when values of $%
\Omega _{c}$ are increased; however, later no major change has been
observed. 

\section{Conclusions }

In this paper to study dust acoustic shock waves, we have
Zakharov--Kuznetsov (ZK) equations by employing reductive perturbation
technique for both cases when dust is inactive and dynamic (fast and slow
time scales). When dust is active both electrons and ions are considered to
follow kappa/Cairns distribution. Main conclusion is that the
superthermality and viscosity in both cases when dust is in inactive and
active, plays as a key role in dissipation for the propagation of acoustic
shock waves. Also charge number ($Z_{d}$) affects the nonlinear coefficient (%
$A^{\prime }$) such as increasing the $Z_{d}$ from $20$ to $72$, the
nonlinear coefficient decreases whereas after 72 it rises abruptly. We would
like to add that charge fluctuations also plays a significant role in the
formation of shocks in dusty plasmas as has been reported earlier\cite{15},
in future we plan to incorporate dust variable charge  into this model to
understand if charge fluctuations play a dominant role,  the viscosity or
superthermality; however, due to complexity of the problem this is propsoed
for the future sequal of this paper. \ Magnetic field and rotation are
responsible for the dispersiveness of the shock weaves. To the best of the
authors' knowledge, this stationary shock solution has not been studied for
the non-Maxwellian rotating viscous dusty plasma system and we believe that
present findings will be useful for the experiments being established to
study rotating dusty plasmas as well as for the PK-4.

$\mathbf{Acknowledgements}$

One of us (Z. E) is grateful to CAAD office, NCP-Islamabd for the
hospitality where this work was formulated. M. M. A is grateful to Zahida
Ehsan for for the hospitality at SPAR CUI Lahore.

Fig. (1): Phase velocity versus kappa ($\kappa )$.

Fig. (2): Phase velocity versus $\alpha $ for the fixed flow speed for fast
mode.

Fig. (3): Phase velocity versus $\alpha $ for the fixed flow speed for slow
mode.

Fig. (4): Phase velocity versus flow speed.

Fig.(5): Nonlinear coefficient $A^{\prime }$ versus $\alpha .$

Fig. (6): $B^{^{\prime }}$ versus parameter $\sigma _{d}=T_{d}/T_{e}Z_{d}.$

Fig. (7): Nonlinear coefficient $A^{\prime }$ versus $Z_{d}.$

Fig. (8): Potential $\phi _{1}$ versus kinematic viscosity coefficient for
the fixed value of $\kappa (=3).$

Fig. (9): For the fixed value of kinematic viscosity, $\phi _{1}$ versus
different value of $\kappa .$

Fig. (10): Potential $\phi _{1}$ versus kinematic viscosity for the fixed
value of $\alpha =0.15.$

Fig. (11): Potential $\phi _{1}$ versus rotation ($\Omega _{c}$) for the
fixed value of $\kappa (=3)$ and $\eta _{o}=0.1.$

Fig. (12): Potential $\phi _{1}$ versus rotation ($\Omega _{c}$) for the
fixed value of $\eta _{o}=0.1$ and $\alpha (=0.45).$

\end{document}